\documentclass[aps,
prb,
reprint,showpacs,amsmath,amssymb,
groupedaddress
]{revtex4-1}

\usepackage{graphicx,tikz,times,mathrsfs}
\pdfoutput=1

\begin{document}

\title{Acoustic Metal}

\author{Mengyao Xie}
\author{Min Yang}
\email{Correspondence email: min@metacoust.com}
\author{Songwen Xiao}
\author{Yunfei Xu}
\author{Shuyu Chen}
\email{Correspondence email: chen@metacoust.com}
\affiliation{
	Acoustic Metamaterials Group Ltd. Genesis, 33-35 Wong Chuk Hang Rd. Hong Kong, China
}

\begin{abstract}
	
Metal reflects electromagnetic waves because of the large conductivity
that is responsible for dissipation. During which the waves undergo a
180$^{\circ}$ phase change that is independent of the frequency.
There is no counterpart material for acoustic waves.  Here we show
that by using an array of acoustic resonators with a designed
high-density dissipative component, an ``acoustic metal'' can be
realised that strongly couples with sound over a wide frequency range
not otherwise attainable by conventional means.  In particular, we
show the acoustic Faraday cage effect that when used as a ring
covering an air duct, 99\% of the noise can be blocked without
impeding the airflow.  We further delineate the underlying volume
requirement for an acoustic metal based on the constraint of the
causality principle.  Our findings complement the missing properties
of acoustic materials and pave the way to the strong wave-material
couplings that are critical for the applications as high-performance
audio devices.

\end{abstract}

\maketitle

\section{Introduction}
\label{sec:introduction}

Large dissipation coefficient does not always lead to effective energy
absorption.  The best example is metal, which has a large conductivity
but reflects the electromagnetic (EM) waves strongly
\cite{hagen1903beziehungen}.  Such reflections are noted to be nearly
180$^{\circ}$ out of phase with the incident wave and independent of
the frequency.  While in recent years resonance-based acoustic
metamaterials \cite{cummer2016controlling, ma2016acoustic,
kadic20193d, assouar2018acoustic} have achieved exceptional material
properties, including zero and negative parameters
\cite{liu2000locally, fang2006ultrasonic, lee2010composite,
nguyen2010total, yang2013coupled, xie2013measurement,
kaina2015negative}, chiral micropolar behaviour
\cite{frenzel2017three, rueger2018strong, zhu2018observation},
nonreciprocity by a magnetic-like circular flow \cite{fleury2014sound,
auregan2017p, fleury2015subwavelength}, and exotic absorptions
\cite{yang2017sound}, there is still no ``acoustic metal'' that
reflects sound in the same way as metal reflects EM waves.  This is
due to the very high contrast between the solid and air impedance that
prevents the effective coupling and damping in broadband.  Recently,
by a designed integration of multiple acoustic resonators, broadband
near-perfect absorption was achieved \cite{yang2017optimal,
zhang2016three, li2018broadband, peng2018composite,
chang2018broadband, zhu2019broadband, liu2019thin, long2019broadband,
huang2020compact}.  In this work, we push the dissipative component of
the resonator array an order of magnitude higher so that the structure
becomes an acoustic metal that reflects sound waves instead of
absorbing them.  In analogy to the metallic mesh that shields the
microwaves from leaking out of the microwave oven or the Faraday cage
for shielding the EM waves, a circular ring of such acoustic metal,
lining the inner wall of a short air duct ($\sim4$ cm), is shown to
block almost all the low-frequency noise within an octave.  Such
blocking effect evidences the strong couplings with sound.

In what follows, Section~\ref{sec:acoust_metal_and_the_soft_boundary}
presents a ``phase diagram'' to position the acoustic metal in
relation to other acoustic materials as characterised by their
impedance.  In Section~\ref{sec:design_strategy} we describe the
design strategy for the acoustic metal, followed by the presentation
of experimental results in Section~\ref{sec:experimental_realisation}.
The underlying volume requirement for an acoustic metal, in relation
to its effective frequency range, is described in
Section~\ref{sec:volume_requirement}.  We conclude with a short
recapitulation in Section~\ref{sec:conclusion}.

\section{Acoustic metal and the soft boundary}
\label{sec:acoust_metal_and_the_soft_boundary}

\begin{figure}
	\centering
	\includegraphics[width=8.6cm]{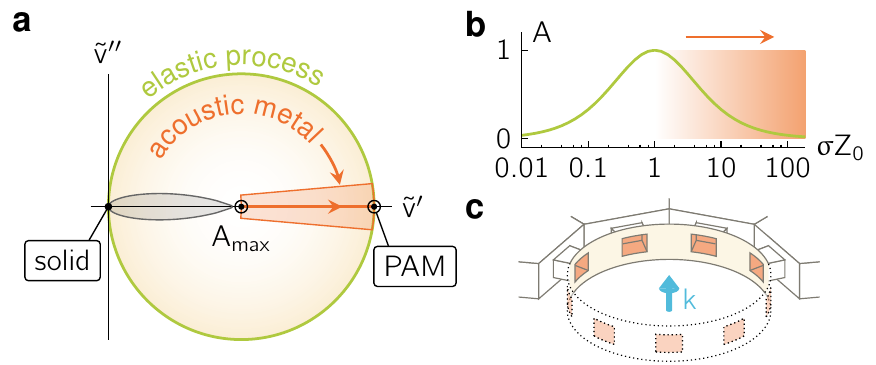}
	\caption{\label{fig:acoustic_metal}
	Acoustic metal.  {\bf a},  Normalised by the displacement velocity
	of the incident sound $v_i$, $\tilde{v}=v/v_i$ of all the different
	sample responses forms a circle on the complex plane
	$\tilde{v}=\tilde{v}'+i\tilde{v}''$.  Most of the solid materials
	are around the origin with the acoustic foam extending to the centre
	(the grey region). In contrast, acoustic metals are in the red
	shaded section in which the reflection is 180 degrees out of phase
	with the incident wave.  Perfect acoustic metal (PAM) is at the
	limit of $\tilde{v}=2$.  The colour gradient maps the relevant
	absorption coefficient.  At the green-coloured circumference the
	absorption is zero.  Absorption is maximised at the centre of the
	circle.  {\bf b}, The absorption as a function of the acoustic
	conductivity $\sigma$ (normalised to the inverse of the specific air
	impedance $Z_0$), plotted along the real axis in {\bf a}.
	$A_\text{max}$ is at $\sigma Z_0=1$.  Acoustic conductivity
	$\sigma\to\infty$ for PAM.  {\bf c}, Like the metallic mesh
	shielding the EM waves, the acoustic metals (the structured
	cavities) covering the inner wall of an air duct (the circular
	cylinder) can block the sound, indicated here by the blue arrow,
	without impeding the airflow.
	}
\end{figure}

By denoting the specific acoustic impedance of the air by $Z_0$, the
displacement velocity on sample's surface equals to the superposition
of the particle velocities of the incident and reflected waves,
$v=p_i/Z_0-p_r/Z_0$. Here, $p_i$ and $p_r$ are the respective incident
and reflected sound wave pressure modulations and the negative sign is
due to their opposite propagating directions.  Conservation of energy
dictates that $|p_r|=|p_i-vZ_0|\leq|p_i|$ which corresponds to the
dimensionless $\tilde{v}=vZ_0/p_i$ inside a circle of radius $1$ on
the complex plane $\tilde{v}=\tilde{v}'+i\tilde{v}''$,
\begin{equation*}
	\left|\tilde{v}-1\right|\leq1,
\end{equation*}
as shown in Figure~\ref{fig:acoustic_metal}a.  This circle can serve
as a ``phase-diagram'' to represent the distinct motions of the
various sample materials in response to the incident sound wave.  At
the origin is the motionless hard boundary.  Along the circular edge,
coloured by green, the total pressure
$p=p_i+p_r=2p_i-vZ_0=(2-\tilde{v})p_i$ always differs from $v$ by a
phase of $\pi/2$, hence lossless.  At the far right of the circle, $p$
approaches zero for the condition of soft boundary, at which point
$\tilde{v}=2$ reaches its maximum magnitude implying the maximum
coupling with sound.  From the circumference of the circle the
absorption coefficient
\begin{equation}
	A=1-\left|{p_r}/{p_i}\right|^2
	=1-\left|\tilde{v}-1\right|^2
	\label{eq:absorption}
\end{equation} 
monotonously increases (indicated by the colour gradient) as a
function of decreasing radial distance to the centre and reaches the
maximum absorption $A_\text{max}$ at $\tilde{v}=1$ (i.e., the
impedance-matching condition).  Most of the solid materials are around
the origin while the porous acoustic materials such as foam and wool
extend to the centre of the circle along the real axis (the grey
region) \cite{yang2017sound}.

Since a metal is characterised by its high conductivity for
dissipation, $\sigma$, hence in analogy we introduce the acoustic
conductivity and permittivity as $v=(\sigma-i\omega\varepsilon)p$.
\cite{jackson1999classical}  Under the incident sound with pressure
modulation $p_i$, $p=2p_i-vZ_0$ so that the dimensionless
\begin{equation}
	\tilde{v}=\frac{2Z_0}{1/(\sigma-i\omega\varepsilon)+Z_0},
	\label{eq:velocity}
\end{equation}
which is along the real axis in Figure~\ref{fig:acoustic_metal}a,
provided $\omega\varepsilon\ll\sigma$ is negligible as that in a
metal.  Furthermore, to have the reflection, $p_r$, 180 degrees out of
phase with $p_i$, where $p_r=(1-\tilde{v})p_i=p_i(1-\sigma
Z_0)/(1+\sigma Z_0)$, $\sigma$ also has to meet another requirement
that $\sigma>1/Z_0$.  Combining the above two conditions, we can
delineate the section coloured in red in
Figure~\ref{fig:acoustic_metal}a as the region for acoustic metals.
This region is noted to be far beyond the scope of the conventional
acoustic materials coloured by grey.  In
Figure~\ref{fig:acoustic_metal}b we show that the absorption has a
peak when $\sigma=1/Z_0$; with the growth of $\sigma$ the perfect
acoustic metal (PAM) limit is approaching when $\sigma\to\infty$ at
the soft boundary limit of (the dimensionless $\tilde{v}=2$).  In
Figure~\ref{fig:acoustic_metal}c we illustrate the geometry of the
acoustic metal resonator array intended to realise the acoustic
counterpart to the Faraday cage effect for the EM waves.

\section{Design strategy}
\label{sec:design_strategy}

\begin{figure}
	\centering
	\includegraphics[width=8.6cm]{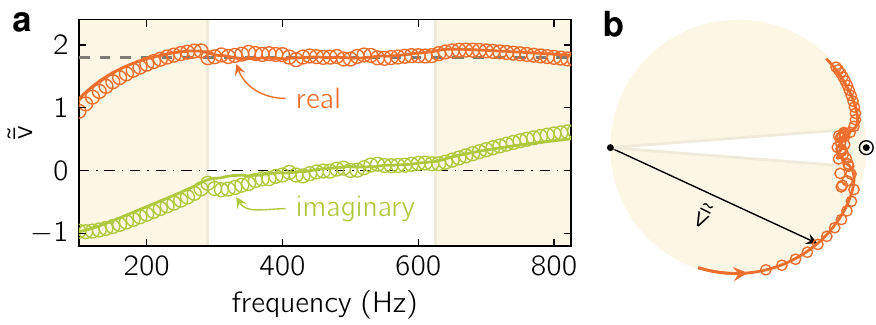}
	\caption{\label{fig:velocity}
		The surface displacement velocity of acoustic metal.  {\bf a}, The
		averaged dimensionless velocity of the acoustic metal,
		$\tilde{\bar v}=\tilde{\bar v}'+i\tilde{\bar v}''$, plotted as a
		function of frequency.   Within the designed frequency band
		(region without shading), $\tilde{\bar v}''\to0$ (green) while
		$\tilde{\bar v}'$ (red) is close to the target values (dashed)
		when $\bar\sigma Z_0=9$ and $\bar\varepsilon=0$.  {\bf b}, On the
		complex plane, the magnitude of $\tilde{\bar v}$ in the designed
		frequency band (region with no shading) are noted to be closely
		distributed around the PAM limit.  The red arrow points to the
		direction of increasing frequency.  In both {\bf a} and {\bf b},
		the solid curves are theoretical predictions from Equation (4) and
		the open circles are retrieved from experimental data of a real
		sample shown in Figure~\ref{fig:HRs}.
	}
\end{figure}

Our strategy is to design an array comprising a large number of
resonators within a sub-wavelength scale so that the resonators
respond in unison.  While the response of each of the resonators can
be delineated mathematically by the Lorentzian form, the array as a
whole would respond as an acoustic metal as shown below.

Consider an array of resonators whose resonance frequencies are
uniformly distributed. The oscillatory nature of the individual
$\varepsilon_n$ may tend to cancel each other while, in contrast, the
individual $\sigma_n$ are always positive and can accumulate.  As a
result, the overall permittivity becomes small while the conductivity
can increase to a large value that is approximately constant in
frequency as desired.

Mathematically, the velocity of the $n$th resonator is given by
\cite{yang2014homogenization} 
\begin{equation}
	v_n=\frac{-i\omega f_n}{\omega_n^2-\omega^2-i\omega\beta_n}p
	=(\sigma_n-i\omega\varepsilon_n)p
	\label{eq:resonator_velocity}
\end{equation}
with $\omega_n$($f_n$) being its resonant frequency(oscillator
strength), and $\beta_n$ the damping coefficient.  Denoting its
surface area by $a_n$, only the averaged
$\bar{v}={\sum_na_nv_n}/{\sum_na_n}$ over all the resonators can
couple to the propagating waves, owing to the sub-wavelength nature of
the array.  Hence $p=2p_i-\bar{v}Z_0$ in
Equation~\eqref{eq:resonator_velocity}.  Together with the definition
of $\bar v$, we obtain an expression similar to
Equation~\eqref{eq:velocity}:
\begin{equation}
	\tilde{\bar v}=\frac{\bar{v}}{p_i/Z_0}
	=\frac{2Z_0}{1/(\bar\sigma-i\omega\bar\varepsilon)+Z_0}.
	\label{eq:average_velocity}
\end{equation}
In Equation~\eqref{eq:average_velocity} the averaged $\bar\sigma$ and
$\bar\varepsilon$ of the composite are defined by
\begin{subequations}
\begin{align}
	\bar\sigma&=\frac{\sum_na_n\sigma_n}{\sum_na_n}=\frac{\omega^2}{\sum_na_n}
	\sum_n\frac{a_nf_n\beta_n}{(\omega_n^2-\omega^2)^2+\omega^2\beta_n^2},
	\label{eq:conductivity}\\
	\bar\varepsilon&=\frac{\sum_na_n\varepsilon_n}{\sum_na_n}=\frac{1}{\sum_na_n}
	\sum_n\frac{a_nf_n(\omega_n^2-\omega^2)}{(\omega_n^2-\omega^2)^2+\omega^2\beta_n^2}.
	\label{eq:permittivity}
\end{align}
\label{eq:susceptibility}
\end{subequations}

\begin{figure*}
	\centering
	\includegraphics[width=17.9cm]{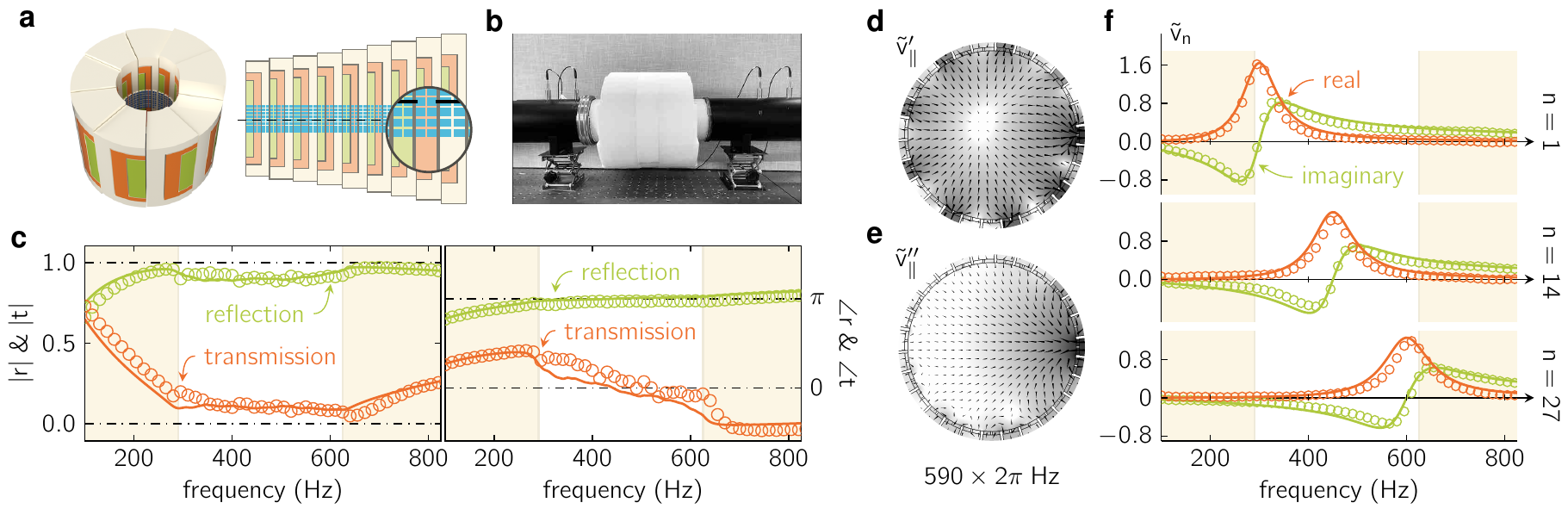}
	\caption{\label{fig:HRs}
		The acoustic metal structure comprises an array of Helmholtz
		resonators (HRs).  {\bf a}, 27 HRs formed a compact structure
		lining the inner wall of an air duct.  The projected pattern of
		their cavities and necks (the magnified blue rectangles) is shown
		by the 2D schematics.  {\bf b}, A photo image of the sample and
		relevant experiment setups.  {\bf c} The amplitude and phase of
		the reflectance $r$ and transmittance $t$.  Within the designed
		frequency band (the region without shading), the transmitted
		energy is only about $|t|^2\sim1\%$ with the reflection phase
		around $\angle r\sim\pi$.  {\bf d} and {\bf e}, The simulated,
		normalised particles' in-plane velocity fields, $\tilde
		v_\|=\tilde v_\|'+i\tilde v_\|''$, plotted on the cross-section
		marked in {\bf a} by the dashed line.  $\tilde v_\|'$ and $\tilde
		v_\|''$ have distinct topology features in which
		$\nabla\cdot\tilde v_\|'\neq0$ indicates the out-plane couplings
		caused by the HRs.  {\bf f}, The $\tilde{v}_n$ of the individual
		HRs when the other 26 were closed.  Three examples have been
		chosen as $n=1,14,27$.  All the solid curves are the theoretically
		calculated design targets and the open circles are the data from
		experiments.
	}
\end{figure*}

Equation~\eqref{eq:conductivity} shows that indeed, the averaged
$\bar\sigma$ can accumulate and become large in magnitude whereas the
summed terms in Equation~\eqref{eq:permittivity} are always opposite
in sign between adjacent $\omega_n$, hence cancelling each other.

For a uniformly distributed $\omega_n$ each separated from the
neighbours by a small $\delta\omega=\omega_n-\omega_{n-1}$, we wish to
focus on the two parameter $f_n$ and $\beta_n$ that are crucial to the
acoustic metal design.  In the limit of $\delta\omega\to0$, the
summation in Equation~\eqref{eq:conductivity} becomes an integral that
gives $f_n\simeq\bar\sigma\times2\delta\omega\sum_na_n/(\pi a_n)$.
\footnote{For $\delta\omega\to0$, the summation in
Equation~\eqref{eq:conductivity} becomes an integral similar to that
in Equation (3b) of Ref.~\onlinecite{yang2017optimal} whose result,
in which $D(\omega)=1/\delta\omega$ and $Z_\text{tar}=1/\bar\sigma$,
is the basis of our design.}  Since large $\bar\sigma$ of acoustic
metal requires large displacement velocity, we show in
Appendix~\ref{sec:supplementary_note_for_the_design_methodology} that,
in contrast to the displacement velocity of a single Lorentz resonator
which is suppressed by the damping, the collective summation of the
array resonators' dissipative component can lead to the overall
$\tilde{\bar v}$ increases with the damping as long as $\beta_n$ does
not exceed $2\delta\omega$.  

In the following, we design the acoustic metal by targeting
$\bar\sigma=9/Z_0$ over a broad frequency band from 290 to 625 Hz.  If
there are 27 resonators, $\delta\omega=12.9\times2\pi$ Hz and
$f_nZ_0\times a_n/\sum_na_n=464.4$ Hz.  Therefore, by choosing
$\beta_n=2\delta\omega$, as shown in Figure~\ref{fig:velocity}a by the
solid curves, Equation~\eqref{eq:average_velocity} predicted that
$\tilde{\bar v}'\gg\tilde{\bar v}''$ within the designed frequency
band (the blank region) with the value of $\tilde{\bar v}'$ close to
the target values (dashed line) when $\bar\sigma=9/Z_0$ and
$\bar\varepsilon=0$.  In the phase-diagram shown in
Figure~\ref{fig:velocity}b, all the vectors of the dimensionless
complex velocities between 290 and 625 Hz are noted to fall into a
sector that has no shading.  It is clear that they were mostly in the
acoustic metal phase as delineated in Figure 1a and tightly
distributed around the PAM limit.

\section{Experimental realisation}
\label{sec:experimental_realisation}

We wish to demonstrate the effect of the acoustic metals by realising
an acoustic counterpart to the Faraday cage effect that blocks EM
waves through a layer of metallic mesh whose pores are smaller than
the wavelengths.  For simplicity, we consider only one pore of a mesh
in practice, designed as a ring covering the inner wall of an air duct
as shown in Figure~\ref{fig:acoustic_metal}c.  The actual duct used in
the demonstration has a cross-sectional area $a=78.5$ $\text{cm}^2$.
The specific impedance for the air in the duct is given by $Z_0=\rho
c\sum_na_n/(2a)$ (detailed in
Appendix~\ref{sec:supplementary_note_for_the_design_methodology}),
where $\rho=1.2$ $\text{kg}/\text{m}^3$ is the air density and $c=343$
$\text{m}/\text{s}$ is the speed of airborne sound.

As shown in Figure~\ref{fig:HRs}a, the acoustic metal sample comprises
27 Helmholtz resonators (HRs) whose cavities share the same depth of
$L=74$ mm and each individual rectangular neck is connected to the
central duct with the same length of $l=2$ mm.  As an approximation,
the resonance frequency of the $n$th HR is given by
$\omega_n=c\sqrt{a_n/(A_nLl)}$ with $A_n=V_n/L$ being the effective
cross-sectional area of the cavity and $V_n$ being the cavity volume.
The frequency $\omega_n$ can be directly tuned by the areal ratio
$a_n/A_n$ so as to be uniformly distributed in the designed frequency
band of (290,625) Hz while the desired values of $f_n$ are obtainable
from adjusting the specific $A_n$.  The detailed values of $A_n/a_n$
and $A_n$ for our design are listed in Table~\ref{tab:parameters-SI}
of Appendix~\ref{sec:supplementary_note_for_the_design_methodology}.
By keeping these areas unchanged, we deformed their shapes so that the
resonators can be assembled compactly.  In order to attain a
sufficiently large $\beta_n$, thin partitions were inserted into the
necks so as to divide each into 16 parts (the tiny blue rectangles in
Figure~\ref{fig:HRs}a), with the purpose of achieving larger air-solid
interfaces, aimed to increase the acoustic dissipation.

We have fabricated the designed structure by the stereolithography 3D
printing technology.  To carry out the measurement, the sample was
connected to two circular impedance tubes of the same inner diameter
on the two sides as shown in Figure~\ref{fig:HRs}b.  During the
experiments, a loudspeaker continuously generates sinusoidal signals
from one end.  Two microphones on the front tube recorded the
reflected sound, $p_r$, from the structure while another two
microphones on the rear measuring the transmitted $p_t$.  Acoustic
foam was put at the rear end to insure there was no reflection.  The
results are shown in Figure~\ref{fig:HRs}c by open circles.  Within
the designed frequency band (the area without shading),
$|t|=|p_t/p_i|\sim0.1$ (red) and $|r|=|p_r/p_i|\sim0.89$ (green)
indicating that about 99\% of the forward propagating energy has been
blocked by the acoustic metal.  The phase of reflection $\angle
r\simeq\pi$ in the designed frequency band that consistent with the
signature of a metal.  Based on these experimental data, we can
retrieve the values of $\tilde{\bar v}$ as shown by the open circles
in Figure~\ref{fig:velocity}.  Excellent agreement with the
theoretical design target (solid curves) is seen.  The relationship
between $\tilde{\bar v}$, $r$ and $t$ is detailed in
Appendix~\ref{sec:supplementary_note_for_the_design_methodology}.

Figures~\ref{fig:HRs}d and e show the numerically simulated in-plane
air velocity field $\tilde v_\|=v_\|/v_i$ on the duct cross-section,
which intersects all the HRs' openings as denoted by the dashed line
in Figure~\ref{fig:HRs}a.  The two figures demonstrate in detail how
the HRs attain their metallic behaviour by blocking the sound waves in
the duct.  For sound frequencies in the designed frequency band, e.g.,
at $\omega=590\times2\pi$ Hz, all the in-phase components, i.e. real
parts, of the displacement velocity ($\tilde v'$) vectors in
Figure~\ref{fig:HRs}d are pointed towards the HRs, with a ``source''
of outward arrows at the centre.  Such a source draws the air in the
ducts from both sides of the plane so that, on the transmission side,
the air displacement velocity is opposite to that of the incident
wave's $v_i$, thereby cancelling it.  Meanwhile, the out-of-phase
component of the displacement velocity ($\tilde v''$) vectors in
Figure~\ref{fig:HRs}e presents a distinct feature that
$\nabla\cdot\tilde v_\|''\simeq0$ everywhere---which means that the
air expelled from the resonating HR is completely compensated by the
intakes of the others with different resonance frequencies.  Such
in-phase and out-of-phase behaviours of the displacement velocity
correspond exactly to the two equations in
Equation~\eqref{eq:susceptibility}.

Be noted that our simulations did not consider the airflow in the air
duct since the airflow speed is usually only a few percent of the
sound speed (wind tunnel being the exception), therefore only has a
negligible effect.

The above conclusions can be further confirmed experimentally by
individually measuring the 27 HRs by blocking the 26 resonators and
retrieving the relevant normalised velocity $\tilde v_n=v_n/v_i$ for
the one not blocked.  As illustrated by the three examples in
Figure~\ref{fig:HRs}f, all the $\tilde v_n'>0$ (denoted by the red
curves) confirms that the velocity is directed towards the relevant
HRs in Figure~\ref{fig:HRs}d.  In contrast, at any given frequency in
the design band, all the HRs at those frequencies below the given
frequency contribute positively to $\tilde{v}_n''$ (green curves), and
those at frequencies above the given frequency contributes negatively
to $\tilde{v}_n''$.  Such behaviour corresponds exactly to that
indicated by Figure~\ref{fig:HRs}e.

\section{Volume requirement}
\label{sec:volume_requirement}

\begin{figure}
	\centering
	\includegraphics[width=8.6cm]{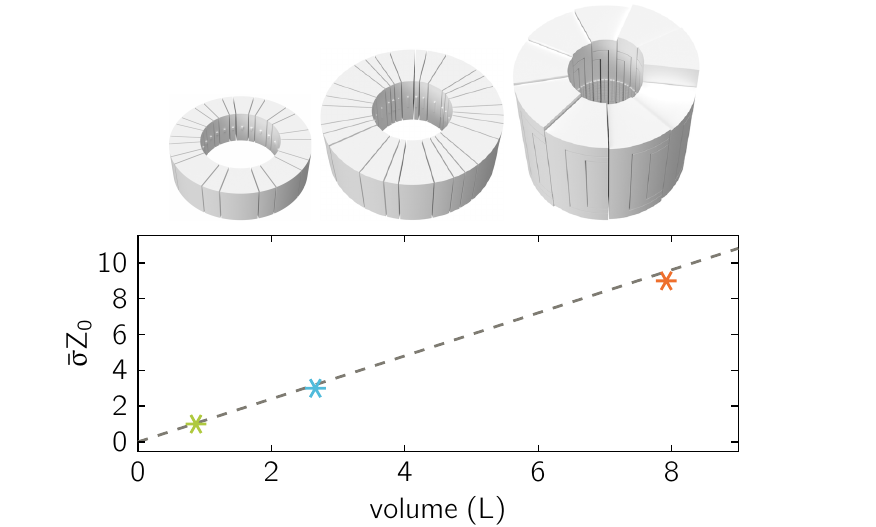}
	\caption{\label{fig:volume_requirement}
		Required volume for an acoustic metal.  According to
		Equation~\eqref{eq:causality}, the principle of causality dictates
		a minimum volume for an acoustic metal that is proportional to its
		target $\bar\sigma Z_0$ as shown by the dashed line.  We have
		verified the relation by checking the volumes (stars) of three
		different experimental samples targeting $\bar\sigma Z_0=1,3,9$ in
		the same frequency range of (290,625) Hz.
	}
\end{figure}

The conductivity, $\bar\sigma$, of an acoustic metal cannot be
arbitrarily large.  As proven in Appendix
\ref{sec:volume_requirement_from_causality_principle}, it is
constrained by the given volume because of the principle of causality
\cite{yang2017sound}.  To be specific, for any structure targeting a
constant $\bar\sigma$ over a broadband frequency regime, causality
dictates that its volume, $V$, must have a minimum volume requirement:
\begin{equation}
	V\geq\bar\sigma Z_0\frac{a}{\pi^2}
	\int_0^\infty\ln\left|\frac{1+\zeta}{1-\zeta}\right|d\lambda=V_\text{min},
	\label{eq:causality}
\end{equation}
where, $\lambda=2\pi c/\omega$ the wavelength and
$\zeta=2i\omega\delta\omega/\pi\times\sum_n1/(\omega_n^2-\omega^2-i\omega\beta_n)$
for the structures of our design.  According to
Equation~\eqref{eq:causality}, $V_\text{min}$ is proportional to
$\bar\sigma Z_0$ as shown in Figure~\ref{fig:volume_requirement} by
the dashed line.  Therefore, a PAM of $\bar\sigma\to\infty$ over a
finite bandwidth will need an infinitely large volume, hence
impossible in practice.  It is worth to mention that the result of
causality constraint in Ref.~\onlinecite{yang2017sound} is a
specific case of Equation~\eqref{eq:causality} with $\bar\sigma
Z_0=1$ and $2a=\sum_na_n$.

We can experimentally verify Equation~\eqref{eq:causality} by checking
the volumes of different samples.  To do so, we fabricated another two
(the schematics in Figure~\ref{fig:volume_requirement}) with smaller
conductivities: $\bar\sigma=3/Z_0$ and $\bar\sigma=1/Z_0$
(impedance-matching).  If we compare the real volumes, $V$, of all the
three samples (stars) with $V_\text{min}$, it is clear that they are
very close to this causally optimal limit.  Hence, they exhibit the
linear relationship seen in Figure~\ref{fig:volume_requirement}.

A useful approximation of Equation~\eqref{eq:causality} (detailed in
Appendix \ref{sec:volume_requirement_from_causality_principle}) is
given by
\begin{equation}
	V\simeq V_\text{min}\simeq\bar\sigma Z_0\frac{2a}{\pi^2}(\lambda_1-\lambda_N),
	\label{eq:approximated_causality}
\end{equation}
in which, $\lambda_{1(N)}=2\pi c/\omega_{1(N)}$ are the wavelengths at
the lower and upper bounds of the designed frequency band.  Take our
first sample as an example, $\bar\sigma Z_0=9$ within the band
(290,625) Hz, so that $2\bar\sigma Z_0(\lambda_1-\lambda_N)/\pi^2$
gives an effective length about $1.1$ m and, for the duct of
cross-sectional area $a=7.85\times10^{-3}$ $\text{m}^2$, $V\simeq8.64$
L that is quite close to the sample's real volume of $7.92$ L.

\section{Conclusion}
\label{sec:conclusion}

We have realised the acoustic counterpart of EM wave's metallic
behaviour by integrating an array of HR resonators each with a large
dissipative coefficient.  Instead of absorbing the acoustic energy,
the composite array strongly reflects sounds within an octave (from
290 to 625 Hz).  A circular ring of acoustic metal lining the inner
wall of an air duct blocks 99\% of the noise energy within a distance
1/30 of the longest wavelength, without impeding the airflow.  One can
therefore expect its application to effectively reduce noise generated
by fans and heavy machinery (as shown in Appendix
\ref{sec:applications_in_ventilation_noise_reduction} and the
Supplementary demo video).  We further delineate the causality
constraint on absorption \cite{yang2017sound, yang2017optimal} to
acoustic metals.  The proposed novel acoustic material properties
enrich the possibilities for new types of insertion loss panels
\cite{ma2013low, yang2015subwavelength, cheng2015ultra,
zhang2017omnidirectional, wu2018high, ghaffarivardavagh2019ultra,
guang2020acoustic, nguyen2020broadband}.  In particular, features such
as large bandwidth, low spectral dispersion, and high susceptibility
are critical for many acoustic elements such as absorbers, lenses, and
audio devices \cite{degraeve2020metamaterial}.

\begin{acknowledgments}
M.Y. wishes to acknowledge the help by Ping Sheng for the inspiring
discussions and numerous suggestions during the preparation and
revision of the manuscript.  M.Y. and M.X. wish to thank Bin Liang,
Jing Yang, and Caixing Fu for helpful discussions.  M.X. and Y.X. wish
to thank Chui Yan Chan, Hin Chi Wong, Ming To Wong, and Man Ying Liew
for their assistance in sample preparations and testing, and Shenzhen
WeNext Technology for their 3D-printing support.
\end{acknowledgments}

\appendix
\section{Supplementary Note for the Design Methodology}
\label{sec:supplementary_note_for_the_design_methodology}

\begin{figure*}
	\centering
	\includegraphics[width=17.9cm]{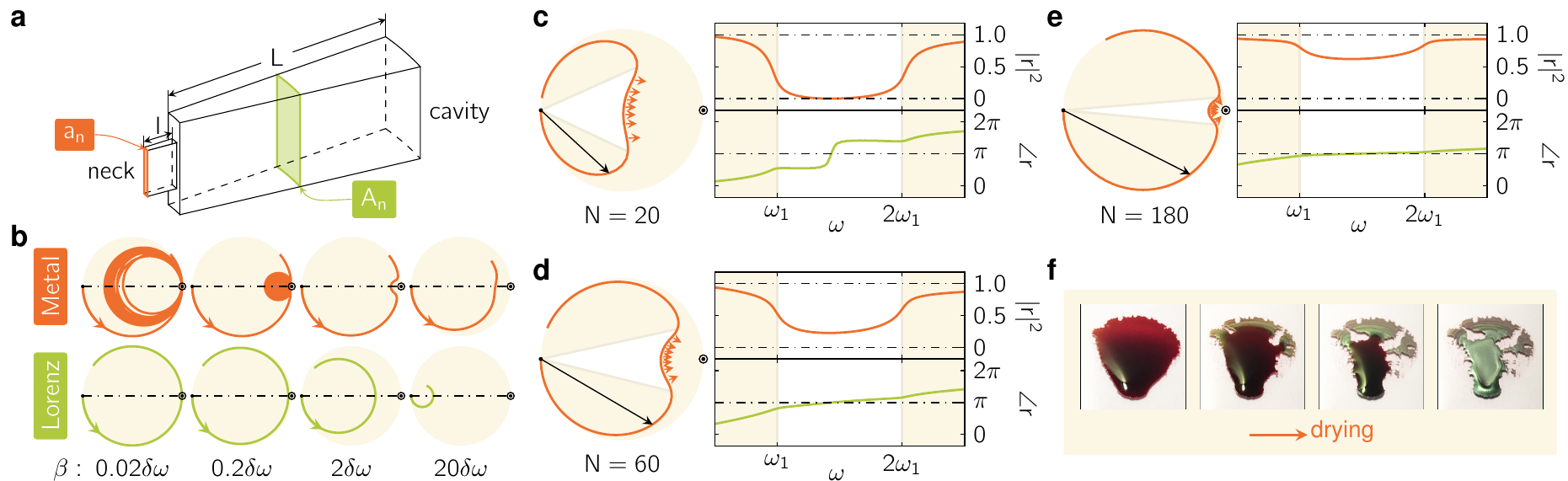}
	\caption{\label{fig:design_details-SI}
		Supplementary to the design methodology. {\bf a}, The schematic
		drawing of a single HR in our acoustic metal.  {\bf b}, Comparison
		between the acoustic metal and a Lorentzian resonator under
		varying system damping coefficients $\beta$.  The curves are the
		trace of their $\tilde{\bar v}$ in the phase-diagram as the
		frequency increases (along the direction denoted by the arrow).
		{\bf c}-{\bf e}, Different absorbers comprise the different number
		of resonators, $N$, distributed in the same frequency range
		$(\omega_1,2\omega_1)$.  On the left columns are their complex
		$\tilde{\bar v}$ on the phase-diagram.  On the right are the
		relevant reflection intensity $|r|^2$ and phase $\angle r$ under
		normal incidents.  The sectors and bands without any shading
		represent the designed frequency band.  {\bf f}, The photo of a
		drop of red ink on glass at different times during its drying.
	}
\end{figure*}

\begin{table}
\begin{tabular}{c|c|c|c|c|c|c|c|c|c}
	\hline
	$n$&	1&	2&	3&	4&	5&	6&	7&	8&	9\\\hline
	$\omega_n/\omega_1$& 1& 1.03& 1.09& 1.12& 1.17& 1.22& 1.28& 1.32& 1.36\\
	$a_n(\text{cm}^2)$& 4.27& 4.48& 4.45& 4.27& 4.13& 4.13& 3.79& 3.79& 3.68\\
	$A_n/a_n$& 18& 16& 16& 15&	14& 14& 13& 13& 12\\
	\hline\hline
	$n$&	10&	11&	12&	13&	14&	15&	16&	17&	18\\\hline
	$\omega_n/\omega_1$& 1.41& 1.46& 1.50& 1.53& 1.57& 1.61& 1.66& 1.70& 1.75\\
	$a_n(\text{cm}^2)$& 3.50& 3.50& 3.44& 3.44& 3.44& 3.38& 3.41& 3.41& 3.26\\
	$A_n/a_n$& 12& 11& 10& 10&	10& 9.5& 9.2& 8.6& 8.7\\
	\hline\hline
	$n$&	19&	20&	21&	22&	23&	24&	25&	26&	27\\\hline
	$\omega_n/\omega_1$& 1.80& 1.84& 1.88& 1.93& 1.97& 2.02& 2.05& 2.10& 2.16\\
	$a_n(\text{cm}^2)$& 3.33& 3.30& 3.28& 3.26& 3.52& 3.54& 4.69& 4.66& 4.62\\
	$A_n/a_n$& 8.2& 7.9& 7.5& 7.3& 6.9& 6.6& 4.5& 4.3& 4.1\\
	\hline
\end{tabular}
	\caption{\label{tab:parameters-SI}
		The resonance frequencies and geometric parameters of the 27 HRs.
		The frequencies are normalised to the resonance of the first HR,
		$\omega_1=290\times2\pi$ Hz and fairly evenly distributed.  $a_n$
		and $A_n$ are the areas of neck and cavity of the $n$th HR
		respectively.
	}
\end{table}

Figure~\ref{fig:design_details-SI}a is a schematic drawing for a
single HR used in our design.  For the $n$th one, we denote the area
of its opening (red) as $a_n$.  If the volume of the
sector-cylindrical cavity is $V_n$ and the depth along the radius is
$L$, we can define its effective area $A_n=V_n/L$ (green).  In our
case, the length of neck is $l=2$ mm and $L=74$ mm, the resonance
frequencies, $\omega_n$, of the 27 HRs together with the associated
$a_n$ and $A_n$ are listed in Table~\ref{tab:parameters-SI}.  Here,
the values of $\omega_n$ were from the FEM simulation by COMSOL
Multiphysics, in which the evanescent waves at the openings and the
weak couplings between neighbouring HRs have all been included.

Unlike the Lorentz resonator whose response was always suppressed by
the damping mechanism, our proposed acoustic metal can have large
damping while simultaneously display large displacement velocity.
Here, we will try to visualise this fact by drawing the $\tilde{\bar
v}$ of our sample in the phase-diagram, as shown by the top row in
Figure~\ref{fig:design_details-SI}b, when the damping coefficient
$\beta$ takes values from $0.02\delta\omega$ to $20\delta\omega$.  It
is clear that the overall motion intensities of $\tilde{\bar v}$
increase if $\beta<2\delta\omega$ and only start to drop when
$\beta>2\delta\omega$.  
%We can understand this result by recalling
%Equation~\eqref{eq:conductivity} and noticing that, since
%$\delta\omega$ is small, for any $\omega$ we can always find an
%$\omega_n\simeq\omega$ and the relevant term dominates the summation.
%By taking $(\omega_n-\omega)^2\simeq\delta\omega^2$,
%\begin{equation}
%	\bar\sigma\simeq\frac{\omega^2}{\sum_na_n}
%	\frac{a_nf_n\beta}{4\omega^2\delta\omega^2+\omega^2\beta^2}
%\end{equation}
%that is increase with $\beta$ when $\beta<2\delta\omega$ and reaches
%maximum when $\beta=2\delta\omega$.  
Therefore, we have chosen the value of $2\delta\omega$ for $\beta$ in
the design.

On the contrary, if we block all the other 26 resonators and leave
only one open (e.g., the 14th one), there is only one term left in the
summation in Equation~\eqref{eq:conductivity} and the relevant
responded velocity has a simple Lorentz-form (as shown in
Figure~\ref{fig:HRs}f),
\begin{equation}
	\tilde{v}_{14}=\frac{-72i\omega\delta\omega a/(\pi\rho ca_{14})}{\omega_{14}^2-\omega^2-i\omega(\beta+18\delta\omega/\pi)}p_i.
\end{equation}
The bottom row of Figure~\ref{fig:design_details-SI}b shows that, in
contrast to what happened in our acoustic metal, its magnitude
continuously shrinks as the growth of $\beta$.

We also wish to show the role of the mode-density in the acoustic
metal by a series of numerical examples.  Let's start from 20
resonators uniformly distributed in an octave, $(\omega_1,2\omega_1)$.
Their $f_n=2\omega_1 a/(19\pi\rho ca_n)$ and
$\beta=2/19\times\omega_1$ so that meet the requirement of a good
broadband absorber.  Figure~\ref{fig:design_details-SI}c shows its
complex $\tilde{\bar v}$ in the phase-diagram together with the
reflected energy ratio $|r|^2$ and phase $\angle r$ under normal
incident sounds.  Nearly perfect absorption, $|r|^2\sim0$, is seen in
the design band but $\angle r$ is away from $\pi$.  Now, if we keep
increasing the number of resonators, $N$, within the same frequency
range, $\tilde{\bar v}$ starts to move along the small red arrows in
the phase-diagram and continue to converge towards the PAM limit, as
illustrated in Figures~\ref{fig:design_details-SI}d and e.
Eventually, $|r|^2\to1$ and $\angle r\to\pi$.

The above phenomenon recalls for us an optical analogy in the drying
of a drop of ink.  As shown in Figure~\ref{fig:design_details-SI}f,
during the process of its drying, a drop of red ink on glass starts to
acquire a metallic greenish reflection, sometimes called ``sheen''.
The reason is that the red ink absorbs out the greens of transmitted
light.  So if the ink is very concentrated, the number of dye
particles per unit volume will be large and can therefore exhibit a
strong surface reflection for the frequencies of green light---similar
as what we have seen in the above simulations for sounds.  That means,
the mechanism in this paper is not limited to acoustics as concluded
by Richard Feynman as ``if \emph{any} material gets to be a
\emph{very} good absorber at any frequency, the waves are strongly
reflected at the surface and very little gets inside to be absorbed.''
\cite{feynman2006lectures}

Finally, let's evaluate $Z_0$ of a narrow duct for the array of
objects on the sidewalls.  By a ``narrow'' duct, we mean one whose
width is small comparing to the wavelengths.  So that only the waves
with uniform front can travel inside.  For the array also small along
the wave direction, the air region in front of it is in sub-wavelength
hence relatively incompressible.  That means the net volume of air in
and out of it is zero.  Mathematically,
$\sum_na_nv_n+2av_\text{rad}=0$ with $v_\text{rad}$ being the particle
displacement velocity of the pair of symmetric radiative sound waves
pointing outward.  For a given objects' motions of $v_n$ that cause a
sound pressure in the front air by $p_\text{rad}$,
$v_\text{rad}=p_\text{rad}/(\rho c)$ so that
$\sum_na_nv_n=-2ap_\text{rad}/(\rho c)$.  Therefore, according to the
definition,
\begin{equation}
	Z_0\equiv-\frac{p_\text{rad}}{\bar v}=\frac{\sum_na_n}{2a}\rho c.
\end{equation}
For a pair of symmetrically incident sound waves in the same amplitude
and phase, the relevant symmetric reflections from the array are given
by
\begin{equation}
	p_r^+=\frac{1/(\bar\sigma-i\omega\bar\varepsilon)-Z_0}{1/(\bar\sigma-i\omega\bar\varepsilon)+Z_0}p_i^+.
\end{equation}
Here, we use the superscript $+$ to denote the symmetric feature.
Meanwhile, anti-symmetric incidents will not excite the motions of the
array, hence the anti-symmetric $p_r^-=-p_i^-$.  

The scenario of sound coming from one side is the superposition of the
symmetric and anti-symmetric scattering processes equally weighted:
\cite{yang2017sound} $p_i=p_i^++p_i^-$ and $p_i^+=p_i^-$.  Therefore,
the transmitted and reflected sounds are given by
\begin{subequations}
\begin{align}
	p_t&=p_r^+-p_r^-=\frac{1/(\bar\sigma-i\omega\bar\varepsilon)}{1/(\bar\sigma-i\omega\bar\varepsilon)+Z_0}p_i,\\
	p_r&=p_r^++p_r^-=\frac{-Z_0}{1/(\bar\sigma-i\omega\bar\varepsilon)+Z_0}p_i,
\end{align}
\end{subequations}
or, in terms of $\tilde{\bar v}$,
\begin{align}
	p_t=\left(1-\frac{\tilde{\bar v}}{2}\right)p_i\text{, and }
	p_r=-\frac{\tilde{\bar v}}{2}p_i.
\end{align}
Because there is always half of the energy in the incoming sound
cannot be coupled to the objects, the maximum of their absorption
coefficient can only be $A_\text{max}=0.5$.

\section{Volume Requirement from Causality Principle}
\label{sec:volume_requirement_from_causality_principle}

\begin{figure}
	\centering
	\includegraphics[width=8.6cm]{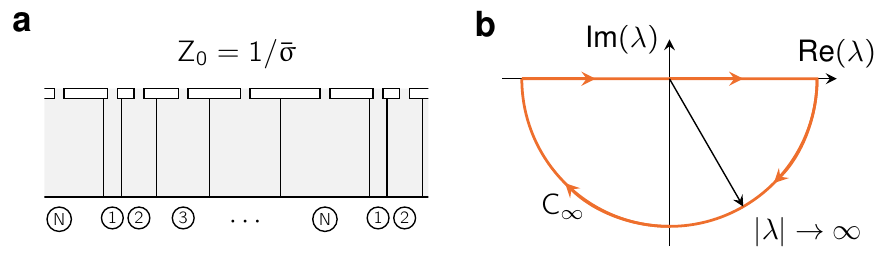}
	\caption{\label{fig:causality-SI}
		{\bf a}, Schematics for the thought experiment to explore the
		causality constraint.  The composite was placed in an imaginary
		medium having the specific impedance $Z_0=1/\bar\sigma$ with
		$\bar\sigma$ being its target conductivity (inside of the cavities
		are still air shown in grey).  {\bf b}, The contour for the
		integral in Equation~\eqref{eq:contour_integral}.
		}
\end{figure}

To explore the role of causality in acoustic metals, we consider a
thought experiment of embedding the sample in an imaginary medium
having the specific impedance $Z_0=1/\bar\sigma$, the reciprocal of
the target acoustic conductivity $\bar\sigma$ (as shown in
Figure~\ref{fig:causality-SI}a).  The reflection process in this
thought experiment should also follow the principle of causality hence
constrained in a similar way as shown in
Ref.~\onlinecite{yang2017optimal}.  That means,
\begin{equation}
	p_{r}(t)=\int_{-\infty}^{\infty}r(\delta t)p_i(t-\delta t)d\delta t.
\end{equation}
with $r(\delta t)=0$ when $\delta t<0$.  Therefore, the reflection
coefficient as a function of frequency,
\begin{equation}
	r(\omega)=\int_{-\infty}^\infty r(\delta t)e^{i\omega\delta t}d\delta t,
\end{equation}
can be analytically continued into the complex frequency and still be
analytic in the upper half-plane.  In terms of the wavelength
$\lambda=2\pi c/\omega$, that means $r(\lambda)$ has no poles in the
lower half-plane of complex $\lambda$ but may have zeros.  Similar as
in Ref.~\onlinecite{yang2017optimal}, we consider the logarithmic
function $\ln r$ which is analytic except in the positions,
$\lambda_n$, where $r=0$.  Such logarithmic singularities can be
removed by multiplying the factor
$(\lambda-\lambda_n^\ast)/(\lambda-\lambda_n)$ with unity amplitude,
where, $\ast$ denotes the complex conjugation.  Then for
\begin{equation}
	\tilde{r}(\lambda)=r(\lambda)\prod_n\frac{\lambda-\lambda_n^\ast}{\lambda-\lambda_n},
\end{equation}
$\ln\tilde{r}(\lambda)$ is analytic with no singularities in the lower
half-plane.

Therefore, according to the Cauchy's theorem, the integral along the
contour in Figure~\ref{fig:causality-SI}b is zero,
\begin{equation}
	\int_C\ln\tilde{r}d\lambda=0.
	\label{eq:contour_integral}
\end{equation}
Because $|\tilde{r}|=|r|$ and $|r(\lambda)|=|r(-\lambda)|$, the real
part of Equation~\eqref{eq:contour_integral} gives
\begin{align}
	-2&\int_0^\infty\ln|r|d\lambda\nonumber\\
	&=\text{Re}\left[\int_{C_\infty}\left(\ln r+
	\sum_n\ln\frac{\lambda-\lambda_n^\ast}{\lambda-\lambda_n}\right)d\lambda\right],
\end{align}
here $C_\infty$ is an lower half-circle with radius tends to infinity.

In the long wavelength limit of $|\lambda|\to\infty$,
\begin{equation}
	\lim_{|\lambda|\to\infty}\ln\frac{\lambda-\lambda_n^\ast}{\lambda-\lambda_n}
	=2i\frac{\text{Im}(\lambda_n)}{\lambda}
\end{equation}
so that
\begin{equation}
	\int_{C_\infty}\ln\frac{\lambda-\lambda_n^\ast}{\lambda-\lambda_n}d\lambda
	=2\pi\text{Im}(\lambda_n).
\end{equation}
Where, $\int_{C_\infty}d\lambda=\int_0^{-\pi}i\lambda d\gamma$ with
$\lambda=|\lambda|e^{i\gamma}$.  Meanwhile, when $\lambda\to\infty$,
the composite was uniformly compressed and expanded under the external
pressure $p_i+p_{r}$.  According to Hooke's law,
$p_i+p_{r}=\kappa_0\Delta V/V=i\kappa_0\bar{v}\sum_na_n/(\omega V)$
with $\kappa_0$ the static bulk modulus of the composite and $V$ their
total volume.  On the other hand, since
$\bar{v}=(p_i-p_{r})/Z_0=\bar\sigma(p_i-p_{r})$, the reflection
coefficient can be solved as
\begin{align}
	r=\frac{p_{r}}{p_i}
	&=\frac{\kappa_0\bar\sigma\sum_na_n+i\omega V}{\kappa_0\bar\sigma\sum_na_n-i\omega V}
	=1+i\frac{1}{\kappa_0}\frac{2V}{\bar\sigma\sum_na_n}\omega+\cdots\nonumber\\
	&=1+i\frac{\kappa}{\kappa_0}\frac{4\pi V}{\bar\sigma\rho c\sum_na_n}\frac{1}{\lambda}+\cdots,
\end{align}
here $\kappa=\rho c^2$ denotes the air bulk modulus.  Hence, $\ln
r\simeq i\kappa/\kappa_0\times4\pi V/(\bar\sigma\rho
c\sum_na_n\lambda)$ at the long wavelength limit and
$\int_{C_\infty}\ln
rd\lambda=\kappa/\kappa_0\times4\pi^2V/(\bar\sigma\rho c\sum_na_n)$.
Therefore,
\begin{align}
	-\int_0^\infty\ln|r|d\lambda
	&=\frac{\kappa}{\kappa_0}\frac{2\pi^2V}{\bar\sigma\rho c\sum_na_n}+\pi\sum_n\text{Im}(\lambda_n)\nonumber\\
	&\leq\frac{\kappa}{\kappa_0}\frac{2\pi^2V}{\bar\sigma\rho c\sum_na_n}
	\label{eq:causality_constraint_1-SI}
\end{align}
since $\text{Im}(\lambda_n)\leq0$.  

For our acoustic metal, according to
Equation~\eqref{eq:average_velocity},
\begin{equation}
	\tilde{\bar v}=2\left[1+\frac{\sum_na_n}{i\omega Z_0}\left(\sum_n\frac{a_nf_n}
	{\omega_n^2-\omega^2-i\omega\beta_n}\right)^{-1}\right]^{-1}
\end{equation}
in the current case, $Z_0=1/\bar\sigma$, and
\begin{align}
	r=1-\tilde{\bar v}=\frac{1+\zeta}{1-\zeta}
\label{eq:reflection-SI}
\end{align}
with
\begin{equation}
	\zeta=\frac{i\omega}{\bar\sigma\sum_na_n}\sum_n\frac{a_nf_n}{\omega_n^2-\omega^2-i\omega\beta_n}.
\end{equation}
Because $f_n=2\bar\sigma\delta\omega\sum_na_n/(\pi a_n)$,
\begin{equation}
	\zeta=i\frac{2\omega\delta\omega}{\pi}\sum_n\frac{1}{\omega_n^2-\omega^2-i\omega\beta_n}
\end{equation}
irrelevant to the target $\bar\sigma$.  Also, if the composite was
rigid cavities filled with air as what in the main text,
$\kappa_0=\kappa$, Equation~\eqref{eq:causality_constraint_1-SI}
becomes
\begin{equation}
	V\geq\bar\sigma\rho c\frac{\sum_na_n}{2\pi^2}\int_0^\infty\ln\left|\frac{1+\zeta}{1-\zeta}\right|d\lambda.
	\label{eq:causality_constraint_2-SI}
\end{equation}
That gives a lower limit for the volume proportional to $\bar\sigma$.

To further evaluate the integral in
Equation~\eqref{eq:causality_constraint_2-SI}, we assume that the
system's averaged acoustic conductivity ideally takes the value of
$\bar\sigma$ within the frequency range $(\omega_1,\omega_N)$ and
being zero elsewhere.  According to the Kramers-Kronig relationship,
\begin{align}
	\bar\sigma-i\omega\bar\varepsilon=-i\frac{2\bar\sigma}{\pi}
	\left[\tanh^{-1}\frac{\omega_1}{\omega}-\tanh^{-1}\frac{\omega_N}{\omega}\right].
	\label{eq:dieletric_constant-SI}
\end{align}
The substitution of Equation~\eqref{eq:dieletric_constant-SI} into
Equation~\eqref{eq:reflection-SI} gives
\begin{equation}
	r=\frac{i\pi/2-\tanh^{-1}(\lambda/\lambda_1)+\tanh^{-1}(\lambda/\lambda_N)}
	{i\pi/2+\tanh^{-1}(\lambda/\lambda_1)-\tanh^{-1}(\lambda/\lambda_N)}
\end{equation}
with $\lambda_{1(N)}=2\pi c/\omega_{1(N)}$.  Its substitution into
Equation~\eqref{eq:causality_constraint_2-SI} gives
\begin{equation}
	V\geq\bar\sigma\rho c\frac{\sum_na_n}{\pi^2}(\lambda_1-\lambda_N).
	\label{eq:causality_constraint_3-SI}
\end{equation}
The above equation can provide an estimate for the volume of an
acoustic metal based only on the knowledge of the target conductivity
and bandwidth.  Usually, this estimate will be a little bit higher
than the volume's actual lower limit, since
Equation~\eqref{eq:conductivity} cannot so perfectly match the target
in practice.

\section{Applications in Ventilation Noise Reduction}
\label{sec:applications_in_ventilation_noise_reduction}

\begin{figure}
	\centering
	\includegraphics[width=8.6cm]{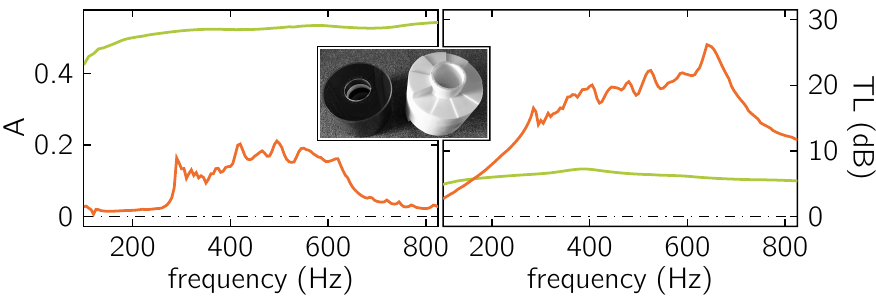}
	\caption{\label{fig:comparison-SI}
		The comparison of absorption coefficient ($A$) and the
		transmission loss (TL) between our composite liner (red) and the
		traditional one of acoustic foam (green).  All the data are from
		the experiments.  The inset is the photo of the two samples.  From
		the two plots it is clear that whereas acoustic foam is mostly an
		absorptive material, our acoustic metal displays superior
		transmission loss with low dissipation.
		}
\end{figure}

The understanding about the causality constraint provides us with a
new perspective for the application of noise control.  According to
Equation~\eqref{eq:causality_constraint_3-SI}, the better blocking
efficiency in an air duct, that corresponds to larger values of
$\bar\sigma$, requires larger volume of $V$.  However, in practice,
there is usually limited volume available for the noise treatments.
In this case, the spectrum design becomes crucial, with which one can
design the reflection ability given by a volume to concentrate in the
frequencies where the noise energy is most intensive, so as to narrow
the bandwidth in Equation~\eqref{eq:causality_constraint_3-SI} for a
larger $\bar\sigma$ of better noise reduction effect.  This strategy
is particularly useful for the noise having a characteristic spectrum,
such as from the fan and machine.

To demonstrate the advantage of our spectrum design, we have compared
our composite liner with a traditional one which comprises an acrylic
cavity filled by acoustic foam, as shown by the inset photo in
Figure~\ref{fig:comparison-SI}.  The volume of the cavity is the same
as our sample and connected to the air duct by the same necks.  As
shown in the right column of Figure~\ref{fig:comparison-SI}, such a
liner exhibited even transmission loss (TL) about 5 (dB) for all the
measuring frequencies.  In contrary, the TL of our liner was maximised
within the design band of $(290,625)$ Hz and presented the advantages
of $10\sim20$ dB.

In acoustics, acoustic foam has been generally considered as a high
dissipative material.  However, as mentioned at the beginning of this
paper, the associated damping is not large enough for being an
acoustic metal.  That is evidenced by the absorption data as shown in
Figure~\ref{fig:comparison-SI}.  The absorption coefficient
$A=1-|t|^2-|r|^2\sim0.5$ indicated that the relevant effective
acoustic conductance was only $\bar\sigma Z_0\simeq1$.  Which is far
below the $\bar\sigma Z_0=9$ of our sample.

To show the sound blocking effect of acoustic metal, we have made a
demo video in the Supplementary Material in which, a Bluetooth
speaker, playing an audio recording of broadband wind noise, was
placed at the bottom of the air duct with the HRs' openings closed by
a metallic ring.  The loud noise can be heard at the top port.  A
Decibel meter near the top port recorded a sound pressure level (SPL)
at 95 dB.  After removing the metallic ring to expose the HRs to the
noises, the emitted SPL suddenly drops to 70 dB.  This contrast
evidenced a very significant 25 dB of the TL for the specific wind
noise.  

\bibliography{References.bib}

\end{document}